\documentclass[11pt]{article}
\usepackage{graphicx}
\usepackage{cs12,html,hyperref}

\begin{document}
 
\title{Confirming the Metal-Rich Nature of Stars with Giant Planets}
 
\author{Nuno C. Santos\altaffilmark{1}, Garik Israelian\altaffilmark{2},
Michel Mayor\altaffilmark{1}}
\altaffiltext{1}{Geneva Observatory, Switzerland}
\altaffiltext{2}{Instituto de Astrof{\'\i}sica de Canarias, Spain}

\index{extra-solar planets}
\index{metallicity}
\index{spectroscopy}
\index{$^6$Li/$^7$Li}

\begin{abstract}

With the goal of confirming the metallicity ``excess'' observed in stars with planetary mass 
companions, we have conducted a high-precision spectroscopic study of a ``comparison'' sample of 
dwarfs included in the CORALIE extra-solar planet survey (Santos, Israelian, \& Mayor 2001). 
The targets were chosen following two basic 
criteria: they make part of a limited volume and they do not present the signature of a planetary host 
companion. The spectroscopic analysis, done using the very same technique as previous works on the 
metallicity of stars with planets, permitted a direct and non-biased comparison of the
samples. The results have revealed that metallicity plays an impressive role on the giant planet 
formation. The chemical composition of the molecular cloud is 
probably the key parameter to form giant planets. Some evidences 
exist, however, showing the possibility of accretion of matter in the stellar outer 
convective zone. These conclusions impose serious constraints on the planetary systems 
formation and evolution models. 

\end{abstract}

\section{Introduction}

\vspace{-13.5truecm}

{\scriptsize \hspace*{-.9truecm} To appear in the proceedings of the 
12th Cambridge workshop on ``Cool Stars, Stellar Systems, and the \\
\hspace*{-0.85truecm} Sun'', July 30th -- August 3rd, 2001, Boulder, 
Colorado, USA.}

\vspace{12.5truecm}

Following the discovery 
in 1995 of the planet orbiting 51\,Peg (Mayor \& Queloz 1995), we have witnessed a 
complete revolution in the field of extra-solar planets. Almost 70 other
exo-planets were unveiled since then, and the most striking and interesting result to 
date is the ``simple'' fact that the discovered systems do not have much in common with our 
own Solar System (see e.g. Udry et al. 2001). The new results are showing that planet formation 
is not as simple as we thought. In particular, following the traditional paradigm for planetary 
formation, the currently found exo-planets were not even supposed to exist. The direct implication
of this results is the strong need to reconsider theories of planetary formation
and evolution. 

To achieve this important goal we need observational constraints. These can be found
by looking at the planetary orbital characteristics, like the distribution of eccentricities and 
periods, or to the distribution of planetary 
masses (e.g. Udry et al. 2001; Mayor \& Santos 2001). But further evidences seem to be coming from the 
planet host stars themselves, namely by the fact that stars with planets are particularly metal-rich 
(Gonzalez 1998; Santos, Israelian \& Mayor 2001 -- hereafter SIM01). 

The fact that stars with planets\footnote{Here we are referring 
to the now known extra-solar planetary host stars, orbited by Jupiter-like planets in relatively 
short period orbits, when compared to the giant planets in our Solar 
System.} are particularly metal-rich has been shaded until recently in one point: 
to compare the metallicity of stars with planets with the metallicity of stars 
``without'' planets, authors were restricted to published metallicity studies of volume 
limited samples of dwarfs in the solar neighborhood (mainly the one from Favata et al. 
1997). This has a few inconvenients,
the most important being that the metallicities for the ``star-with-planet'' 
and the Favata et al. sample were determined using different sources for the 
atmospheric parameters (spectroscopic vs. colours) -- see e.g. Santos, 
Israelian \& Mayor (2000) -- hereafter SIM00. 
This may introduce systematic errors, and one could expect that 
the difference between the two samples was simply reflecting a bias.

\begin{table}[t]
\begin{center}
\caption{Derived abundances and atmospheric parameters for the 7 planet-hosts not included in the study of SIM01.}
\begin{tabular}{lcccrr}
\noalign{\smallskip}
\tableline
\multicolumn{1}{l}{HD} & \multicolumn{1}{c}{$T_{eff}$} & $\log{g}$      & $\xi_t$        & [Fe/H] & Reference\\
\multicolumn{1}{l}{Number} & \multicolumn{1}{c}{(K)} & (cm\,s$^{-2}$)      & (km\,s$^{-1}$)        &  & \\
\tableline 

HD\,106252    &    5890  &  4.40  &  1.06  &  $-$0.01 & This paper\\
HD\,141937    &    5925  &  4.62  &  1.16  &  0.11    & This paper\\
HD\,160691    &    5820  &  4.44  &  1.23  &  0.33    & This paper\\
HD\,178911\,B &    5650  &  4.65  &  0.85  &  0.28    & Zucker et al. (2001)\\ 
HD\,179949    &    6235  &  4.41  &  1.38  &  0.21    & This paper\\
HD\,195019    &    5830  &  4.34  &  1.24  &  0.09    & This paper\\
HD\,213240    &    5975  &  4.32  &  1.30  &  0.16    & Santos et al. (2001)\\

\noalign{\smallskip}
\tableline 
\end{tabular}
\end{center}
\end{table}

With the goal of settling down the question about the high 
metallicity content of stars with planets, we have recently conducted (SIM01) a spectroscopic study 
of a volume limited sample of 43 stars included in the CORALIE
planet search programme (Udry et al. 2000)\footnote{See also http://obswww.unige.ch/$\sim$udry/planet/planet.html}, and for which the radial-velocity 
seems to be constant over a large time interval. 
In SIM01 we have shown that the currently known stars with giant planets are in average more metal-rich 
than ``field stars'' for which 
there is no radial-velocity signature of planets. Furthermore, the results exclude with great significance 
a ``pollution'' scenario. In this paper we will review these results, adding 7 more stars to the
planet-host sample, and showing that the new points do confirm the results presented in SIM01.

\section{The Data}

\index{*HD 106252}
\index{*HD 141937}
\index{*HD 160691}
\index{*HD 178911B}
\index{*HD 179949}
\index{*HD 195019}
\index{*HD 213240}

The technique and analysis, as well as the data, were presented and discussed in SIM00 and SIM01. In the meanwhile
we have obtained spectra for 6 stars with planets not included in previous studies\footnote{Both using the
CORALIE spectrograph, at the 1.2-m Euler Swiss telescope, at La Silla (ESO), Chile, and the UES spectrograph, at the
William Herschel Telescope, at La Palma, Canary Islands.}.
The results of the spectroscopic analysis for those 6 ``new'' objects are presented in the table.
Another planet host (HD\,178911\,B), not analyzed by us but by someone from the group 
(using the same technique, line-lists and model atmospheres) is also included in the table (with the 
correct reference). Errors were discussed in SIM01.

\section{The Metallicity Excess}

In Fig.\,1 (left) we can see
a comparison between the [Fe/H] distribution of our volume limited sample of field dwarfs 
without detected planetary mass companions, and the same distribution for the stars with 
planets. There is a remarkable difference between both distribution, 
as can be seen from their cumulative functions -- Fig.\,1 (right).

\begin{figure}[t]
\hspace{0.5cm}
\includegraphics*[angle=0, width=12cm]{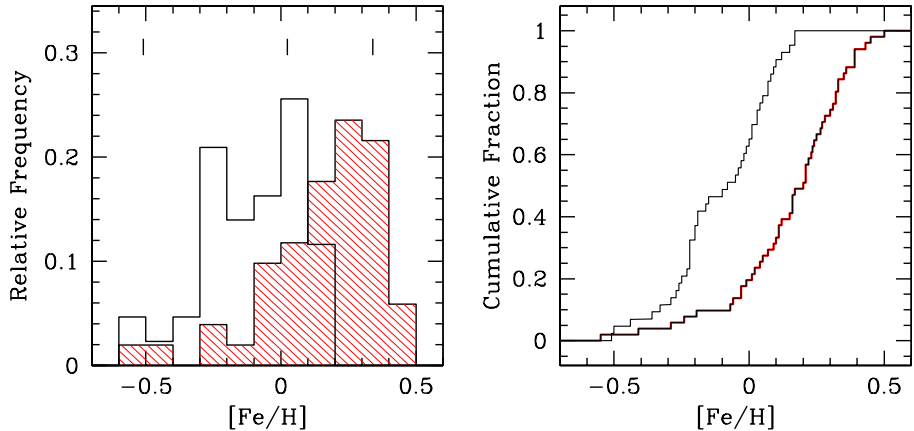}
\caption{{{\it Left}: [Fe/H] distribution of stars with planets (shaded histogram) compared 
with the same distribution of field dwarfs in the solar neighborhood (open histogram). 
The vertical lines represent stars with brown dwarf candidate companions having minimum masses 
between 10 and 20\,M$_\mathrm{Jup}$. {\it Right}: The cumulative functions of both samples. 
A Kolmogorov-Smirnov test shows the probability of the stars' being part of the same sample 
is around 10$^{-8}$.}}
\end{figure}

The plots leave no doubts: stars with planets, or at least with planets similar
to the ones we are finding today, are clearly more metal rich that stars without planetary
companions. Even if we look just at the table, all but one of the 7 planet hosts
with ``new'' [Fe/H] measurements have metallicity higher than solar.
The mean [Fe/H] difference between both samples is around 0.25\,dex. 
As discussed in SIM01, given the uniformity of the analysis and the
absence of observational biases in both samples, these results represent a real trend.

In the figure, stars with low-mass brown-dwarf companions (10\,M$_{\mathrm{Jup}}$ $<$ M$_{2}\,\sin{i}$ $<$ 17\,M$_{\mathrm{Jup}}$) 
are denoted by the vertical lines. No conclusion can be taken at this moment concerning
these cases, but a large dispersion seems to be present.

More interesting conclusions can be taken by looking at the shape of the distribution
of stars with planets alone -- Fig.\,2 (left). As can be seen from the plot, this distribution is rising with [Fe/H], 
up to a value of $\sim$0.4, 
after which we see a sharp cutoff. This cutoff suggests that we may be looking at the approximate 
limit on the metallicity of the stars in the solar neighborhood. On the other hand, the steep rise 
cannot be explained by any sampling bias.

\begin{figure}[t]
\hspace{0.5cm}
\includegraphics*[angle=0, width=12cm]{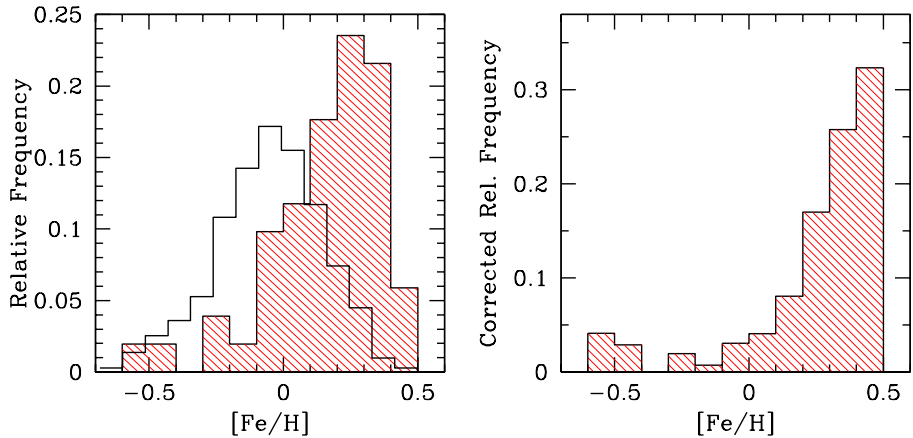}
\caption{{\it Left}: metallicity distribution of stars with planets (dashed histogram) compared 
with the distribution of a large volume limited sample of field dwarfs (empty histogram) -- see SIM01 for more details; {\it right}: correcting the distribution of stars with planets from the same distribution for stars in the volume-limited sample results in a even more steep rise of the planet host star distribution as a function of [Fe/H].}
\end{figure}

The trend is even more clear if we correct the distribution for the fact that the peak
in the metallicity distribution for stars in the solar neighborhood is around solar metallicity:
it is more easy to find a star with [Fe/H]=0 than with [Fe/H]=0.5. The result, presented in Fig.\,2 (right),
leaves no doubts: the probability of finding a planet host is a strong function of its metallicity.
As discussed in SIM01, this cannot be the result of any observational bias,
and is just probably telling us that the probability of forming a giant planet depends strongly on the
metallicity of the gas that gave origin to the star and planetary system.
And although it is unwise to take any strong conclusions based on one point, it is worth of note that
our own Sun is in the ``metal-poor'' tail of the planet hosts [Fe/H] distribution!

Finally, the small ``bump'' seen for low metallicities in the corrected distribution is clearly not 
statistically significant, since only one planet host per bin exists in this region of the plot. Only the addition of more data will permit to say more on this point. If it remains with better statistics, 
or with the discovery of very low mass companions (a few earth masses) it could suggest 
e.g. that the current trend is valid only for the giant planets found today, and that 
very low mass planetary companions can be easily formed even in low metallicity environments. 
Note that the planet around HD\,6434, the lower metallicity planet host in the histograms (SIM01), has a minimum mass below 0.5\,M$_{\mathrm{Jup}}$.

\index{*HD 6434}
\index{*HD 75289}

\section{The ``Primordial'' Origin}

\begin{figure}[t]
\hspace{0.5cm}
\includegraphics*[angle=0, width=10cm]{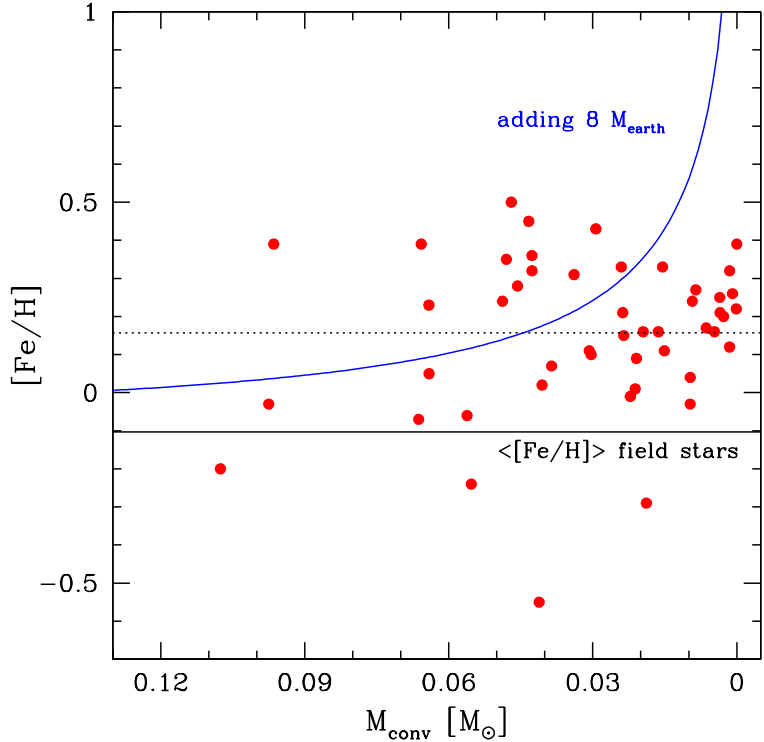}
\caption{Metallicity against convective envelope mass for stars with planets (dots). The [Fe/H]=constant
line represents the mean [Fe/H] for the non-planet hosts stars of Figure\,1. The curved line
represents the result of adding 8 earth masses of iron to the convective envelope of 
stars having an initial metallicity equal to the non-planet hosts mean [Fe/H]. The resulting trend
has no relation with the distribution of the stars with planets.}
\end{figure}

Two different interpretations have been given to the [Fe/H] ``excess'' observed for stars with 
planets. One suggests that the high metal content is the result of the accretion of 
planets and/or planetary material into the star (e.g. Gonzalez 1998). Another, simply states that the 
planetary formation mechanism is dependent on the metallicity of the proto-planetary disk: 
according to the ``traditional'' view, a gas giant planet is formed by runaway accretion 
of gas by a $\sim$10 earth masses planetesimal. The higher the metallicity (and thus 
the number of dust particles) the faster a planetesimal can grow, and the higher the 
probability of forming a giant planet before the gas in the disk dissipates.

There are multiple ways of deciding between the two scenarios (see discussion in
e.g. SIM01). Probably the most clear and strong argument 
is based on stellar internal structure, in particular on the fact that material falling into 
a star's surface would induce a different increase in [Fe/H] depending on the stellar mass, 
i.e. on the depth of its convective envelope (where mixing can occur). However, the data shows no
such trend (see Fig.\,3). In particular, a quick look at the plot indicates that the
upper envelope of the points is quite constant. A similar conclusion was also recently taken 
by Pinsonneault et al. (2001) that showed that even non-standard models of convection and diffusion 
cannot explain the lack of a trend and sustain ``pollution'' as the source of the high-[Fe/H].

Together with the fact that evolved stars with planets also show high-metallicity values, 
and that it seems to be quite difficult to explain the sharp cutoff in the metallicity 
distribution of stars with planets using simple pollution models (SIM01), the facts presented 
here strongly suggest a ``primordial origin'' to the high-metal content of stars with giant planets. This result implies that the metallicity 
is a key parameter controlling planet formation and evolution, and may have enormous 
implications on theoretical models.

\section{Metallicity and Orbital Parameters}

Given that we already have about 55 extra-solar planet hosts with
high-precision and uniform metallicity determinations, we can start to think about
looking for possible trends in [Fe/H] with planetary mass, semi-major axis 
or period, and eccentricity. 

Gonzalez (1998) and Queloz et al. (2000) have shown evidences that 
stars with short-period planets (i.e. small semi-major axes) may be particularly metal-rich,
even amongst the planetary hosts. The number of planets that were known by that time was, 
however, not enough to arrive at a definitive conclusion. 

\begin{figure}[t]
\hspace{0.5cm}
\includegraphics*[angle=0, width=12cm]{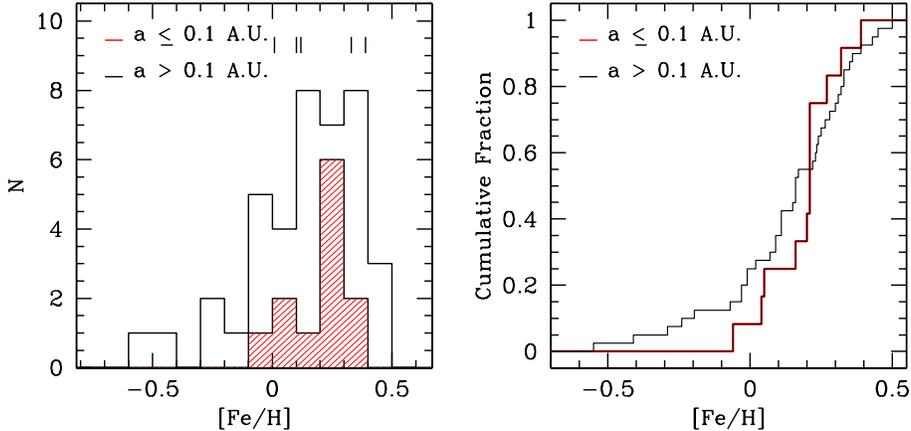}
\caption{{\it Left}: [Fe/H] distributions for planets orbiting at distances shorter and higher 
than 0.1\,AU from their host stars (the hashed and open bars, respectively). {\it Right}: cumulative functions
of both distributions. A Kolmogorov-Smirnov test gives a probability of $\sim$0.7 that both
samples make part of the same distribution.}
\end{figure}

Recent analysis (SIM01) seems in fact to discard any special trend. The result,
also shown here in Fig.\,4 (now including the 7 planet hosts presented in the table and not
 included in the study of SIM01), shows a small tendency for short period systems to
have higher metallicity (see the cumulative functions). However, this tendency is clearly
not significant; the Kolmogorov-Smirnov probability that both samples belong to the same
population is $\sim$0.7. 

In Fig.\,4 the multi-planetary systems, not included in the histograms, are denoted by the
vertical lines\footnote{These include the systems around $\upsilon$\,And 
(Butler et al. 1999), HD\,83443 (Mayor et al. 2001a), 
HD\,168443 (Udry et al. 2001), HD\,82943 (Mayor et al. 2001b), 
and 47\,UMa (Fischer et al. 2001)}. Although we do not have many points, the current results 
also do not show any strong trend (although they all seem to be particularly metal-rich).

\index{*47 UMa}
\index{*HD 83443}
\index{*HD 82943}
\index{*$\upsilon$ And}
\index{*HD 168443}

The same situation can be found concerning other orbital parameters, like the 
eccentricity or planetary mass. However, current results do not discard that e.g. when 
much lighter planets or when systems more similar to the Solar System are found some 
trend may appear (see also discussion in section 3).

\section{HD\,82943: a Case of Pollution}

\index{*HD 82943}

Although the results presented above seem to rule out pollution as the key parameter 
inducing the high metallicity of planet hosts stars, some evidences of pollution have been 
discussed in the literature (e.g. Gonzalez 1998; Laws \& Gonzalez 2001). Perhaps the strongest evidence 
for such phenomena
came recently from the detection of an ``anomalous'' $^6$Li/$^7$Li ratio on the
star HD\,82943, a late F dwarf known to have two orbiting planets (Israelian et al. 2001).

\begin{figure}[t]
\hspace{0.5cm}
\includegraphics*[angle=0, width=11cm]{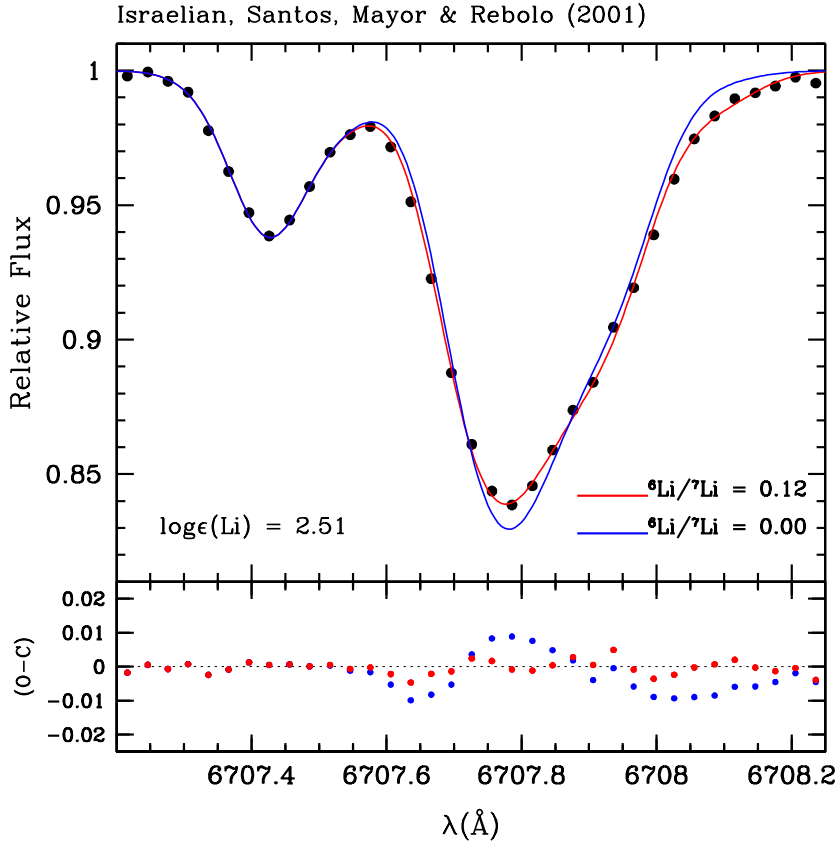}
\caption{$^6$Li signature on the spectrum of HD\,82943 (dots) and two spectral synthesis, one
with no $^6$Li and the other with a $^6$Li/$^7$Li ratio of 0.12 (compatible with the meteoritic value).
The O-C residuals of both fits are shown.}
\end{figure}

The rare $^6$Li isotope represents an unique way of 
looking for traces of ``pollution''. So far, this isotope had been detected in
only a few metal-poor halo and disc stars, but never with a
high level of confidence in any metal-rich or even solar-metallicity
star. Standard models of
stellar evolution predict that $^6$Li nuclei are efficiently destroyed 
during the early evolution of solar-type stars and disappear from their atmospheres 
within a few million years. Planets, however, do not reach
high enough temperatures to burn $^6$Li nuclei, and fully preserve
their primordial content of this isotope. A planet engulfed by its parent
star would boost the star's atmospheric abundance of $^6$Li. 

In fact, planet engulfment following e.g. planet-planet (Rasio \& Ford 1996) or planet-disk 
(Goldreich \& Tremaine 1980) interactions seems to be the only convincing and the less speculative 
way of explaining the presence of this isotope in the atmosphere of HD\,82943. 

It is important to note, however, that the quantity of material we need to add
to the atmosphere of HD\,82943 in order to explain the lithium isotopic ratio
would not be able to change the [Fe/H] of the star by more than a few cents of a dex.  
Furthermore, it it not clear how often this kind of events occur. In this context we 
would like to call attention to the contribution by Garcia Lopez et al. (2001, this book).
In any case, the conclusions presented above, supporting a ``primordial'' source for the high [Fe/H]
of planet host stars are not dependent on this cases of ``pollution''.

Note also that we are referring to the fall of planets or planetary material after the star has reached the main-sequence phase and fully developed a convective envelope; if engulfment happens before that, all planetary material will be deeply 
mixed, and no traces of pollution might be found. It is interesting to note that the whole 
giant-planetary formation phase\footnote{This is probably also true for all the action
concerning hypothetical planet/planetary material engulfment.} must take place when a disk of gas (and debris) is present. Massive gas disks
may not exist at all when a star like the Sun reaches the main-sequence phase (inner 
disks seem to disappear after $\sim$10\,Myr -- e.g. Haisch, Lada, \& Lada -- although recent 
work by Thi et al. 2001 
suggests that gas disks, associated with debris disks, and that are massive enough to form ``jupiters'', may survive up to 30\,Myr).  Thus, all the ``massive'' infall that would be capable of changing
the measured elemental abundances if the star were already at the main-sequence might simply occur
too early, explaining why we do not see strong traces of pollution (in particular, concerning iron).

Although a bit speculative, the fact that no significant ``pollution'' is seen can 
be used as an evidence that disks life-times are in general
shorter than pre-main-sequence duration, if we suppose that planets (in particular planetary 
building blocks) are in fact engulfed by the star at some stage (e.g. Murray et al. 1998).
This fact, if confirmed, could deepen the problem concerning the planetary formation time-scales.

\section{Conclusions}

The results presented in this paper, most of them based and already presented in SIM01 and Israelian et al. (2001), 
can be summarized as follows:

\begin{itemize}

\item The currently known stars with planets are substantially
 metal-rich when compared with non-planetary 
host dwarfs. The mean difference in [Fe/H] is $\sim$0.25\,dex and is clearly significant.

\item The shape of the metallicity distribution of stars with planets has a very clear rise with 
[Fe/H]. This indicates that the probability of finding (and forming) a planet is a strong
function of [Fe/H].

\item ``Pollution'' does not seem to play an important role in determining the high metal content
of the planet host stars. The excess metallicity seems to have a ``primordial'' origin.

\item Some traces of pollution seem to exist concerning light-element abundances, and in
particular the lithium isotopic ratio. The frequency of such events is, however, still not known.

\item An analysis of the planetary orbital parameters (a, m$_2\,\sin{i}$, and e) does not reveal 
any clear trends with [Fe/H].

\end{itemize}

These results can be basically summarized in one sentence: planetary formation and/or evolution, 
or at least the formation of the planetary systems we are currently detecting, seems to be
extremely dependent and sensitive on the metallicity of the cloud that gives origin to the
star/planet system. Given the strong observational constraints this work is giving, it would be
interesting to compare the current results with models in order to better understand exactly 
why this is so.

\acknowledgments

  We wish to thank the Swiss National Science Foundation (Swiss NSF) for
  the continuous support to this project. We also want to thank all the members
  of the CORALIE team, Stephane Udry, Dominique Naef, Didier Queloz, Francesco Pepe,
  as well as Rafael Rebolo from the IAC, without whom this work would not 
  have been possible. Support from Funda\c{c}\~ao para a Ci\^encia e Tecnologia, Portugal, 
  to N.C.S. in the form of a scholarship is gratefully acknowledged.

\end{document}